\documentclass[compress,sort,final,11pt,3p]{elsarticle}

\usepackage[colorlinks]{hyperref}
\usepackage{amsmath,amssymb}
\usepackage{upgreek}
\usepackage{multirow}
\usepackage{tablefootnote}
\usepackage{listings}
\usepackage{xcolor}
\usepackage{booktabs}
\usepackage{xspace}

\usepackage[utf8]{inputenc}

\journal{Computer Physics Communications}
\newcommand\YAMLcolonstyle{\color{black}\ttfamily}
\newcommand\YAMLkeystyle{\color{red!70!black}\ttfamily}
\newcommand\YAMLvaluestyle{\color{green!40!black}\ttfamily}
\newcommand\YAMLliststyle{\color{green!40!black}\ttfamily}

\makeatletter

\newcommand\language@yaml{yaml}

\expandafter\expandafter\expandafter\lstdefinelanguage
\expandafter{\language@yaml}
{
  keywords={true,false,null,y,n},
  keywordstyle=\YAMLvaluestyle,
  basicstyle=\YAMLkeystyle\footnotesize,                                 
  sensitive=false,
  showstringspaces=false,
  comment=[l]{\#},
  morecomment=[s]{/*}{*/},
  commentstyle=\color{purple}\ttfamily,
  stringstyle=\YAMLvaluestyle\ttfamily,
  moredelim=[l][\color{black!50}]{[...},
  moredelim=[l][\color{orange}]{\&},
  moredelim=**[il][\YAMLcolonstyle{:}\YAMLvaluestyle]{:},   
  morestring=[b]',
  morestring=[b]",
  literate =    {---}{{\ProcessThreeDashes}}3
                {>}{{\textcolor{red}\textgreater}}1     
                {|}{{\textcolor{red}\textbar}}1 
                {\ -\ }{{\YAMLliststyle\ {\YAMLcolonstyle-}\ }}3,
  frame=single,
    rulecolor=\color{ipython_frame},
    frame=single,
    frameround={t}{t}{t}{t},
    framexleftmargin=6mm,
    numbers=left,
    numberstyle=\tiny\color{halfgray},
    backgroundcolor=\color{ipython_bg},
    aboveskip=1.5em,
    belowskip=1.5em,
}

\lst@AddToHook{EveryLine}{\ifx\lst@language\language@yaml\YAMLkeystyle\fi}
\makeatother

\definecolor{maroon}{cmyk}{0, 0.87, 0.68, 0.32}
\definecolor{halfgray}{gray}{0.55}
\definecolor{ipython_frame}{RGB}{207, 207, 207}
\definecolor{ipython_bg}{RGB}{247, 247, 247}
\definecolor{ipython_red}{RGB}{186, 33, 33}
\definecolor{ipython_green}{RGB}{0, 128, 0}
\definecolor{ipython_cyan}{RGB}{64, 128, 128}
\definecolor{ipython_purple}{RGB}{170, 34, 255}

\usepackage{listings}
\lstdefinelanguage{iPython}{
    morekeywords={access,and,break,class,continue,def,del,elif,else,except,exec,finally,for,from,global,if,import,in,is,lambda,not,or,pass,print,raise,return,try,while},%
    morekeywords=[2]{abs,all,any,basestring,bin,bool,bytearray,callable,chr,classmethod,cmp,compile,complex,delattr,dict,dir,divmod,enumerate,eval,execfile,file,filter,float,format,frozenset,getattr,globals,hasattr,hash,help,hex,id,input,int,isinstance,issubclass,iter,len,list,locals,long,map,max,memoryview,min,next,object,oct,open,ord,pow,property,range,raw_input,reduce,reload,repr,reversed,round,set,setattr,slice,sorted,staticmethod,str,sum,super,tuple,type,unichr,unicode,vars,xrange,zip,apply,buffer,coerce,intern},%
    sensitive=true,%
    morecomment=[l]\#,%
    morestring=[b]',%
    morestring=[b]",%
    morestring=[s]{'''}{'''},
    morestring=[s]{"""}{"""},
    morestring=[s]{r'}{'},
    morestring=[s]{r"}{"},%
    morestring=[s]{r'''}{'''},%
    morestring=[s]{r"""}{"""},%
    morestring=[s]{u'}{'},
    morestring=[s]{u"}{"},%
    morestring=[s]{u'''}{'''},%
    morestring=[s]{u"""}{"""},%
    literate=
    {á}{{\'a}}1 {é}{{\'e}}1 {í}{{\'i}}1 {ó}{{\'o}}1 {ú}{{\'u}}1
    {Á}{{\'A}}1 {É}{{\'E}}1 {Í}{{\'I}}1 {Ó}{{\'O}}1 {Ú}{{\'U}}1
    {à}{{\`a}}1 {è}{{\`e}}1 {ì}{{\`i}}1 {ò}{{\`o}}1 {ù}{{\`u}}1
    {À}{{\`A}}1 {È}{{\'E}}1 {Ì}{{\`I}}1 {Ò}{{\`O}}1 {Ù}{{\`U}}1
    {ä}{{\"a}}1 {ë}{{\"e}}1 {ï}{{\"i}}1 {ö}{{\"o}}1 {ü}{{\"u}}1
    {Ä}{{\"A}}1 {Ë}{{\"E}}1 {Ï}{{\"I}}1 {Ö}{{\"O}}1 {Ü}{{\"U}}1
    {â}{{\^a}}1 {ê}{{\^e}}1 {î}{{\^i}}1 {ô}{{\^o}}1 {û}{{\^u}}1
    {Â}{{\^A}}1 {Ê}{{\^E}}1 {Î}{{\^I}}1 {Ô}{{\^O}}1 {Û}{{\^U}}1
    {œ}{{\oe}}1 {Œ}{{\OE}}1 {æ}{{\ae}}1 {Æ}{{\AE}}1 {ß}{{\ss}}1
    {ç}{{\c c}}1 {Ç}{{\c C}}1 {ø}{{\o}}1 {å}{{\r a}}1 {Å}{{\r A}}1
    {€}{{\EUR}}1 {£}{{\pounds}}1
    {^}{{{\color{ipython_purple}\^{}}}}1
    {=}{{{\color{ipython_purple}=}}}1
    {+}{{{\color{ipython_purple}+}}}1
    {*}{{{\color{ipython_purple}$^\ast$}}}1
    {/}{{{\color{ipython_purple}/}}}1
    {+=}{{{+=}}}1
    {-=}{{{-=}}}1
    {*=}{{{$^\ast$=}}}1
    {/=}{{{/=}}}1,
    literate=
    *{-}{{{\color{ipython_purple}-}}}1
     {?}{{{\color{ipython_purple}?}}}1,
    identifierstyle=\color{black}\ttfamily,
    commentstyle=\color{ipython_cyan}\ttfamily,
    stringstyle=\color{ipython_red}\ttfamily,
    keepspaces=true,
    showspaces=false,
    showstringspaces=false,
    rulecolor=\color{ipython_frame},
    frame=single,
    frameround={t}{t}{t}{t},
    framexleftmargin=6mm,
    numbers=left,
    numberstyle=\tiny\color{halfgray},
    backgroundcolor=\color{ipython_bg},
    basicstyle=\footnotesize\ttfamily,
    keywordstyle=\color{ipython_green}\ttfamily,
    aboveskip=1.2em,
    belowskip=1.2em,
}

\lstdefinelanguage{MyMathematica}{
    sensitive=true,%
    morecomment=[l](*,%
    morestring=[b]',%
    morestring=[b]",%
    morestring=[s]{'''}{'''},
    morestring=[s]{"""}{"""},
    morestring=[s]{r'}{'},
    morestring=[s]{r"}{"},%
    morestring=[s]{r'''}{'''},%
    morestring=[s]{r"""}{"""},%
    morestring=[s]{u'}{'},
    morestring=[s]{u"}{"},%
    morestring=[s]{u'''}{'''},%
    morestring=[s]{u"""}{"""},%
    literate=
    {á}{{\'a}}1 {é}{{\'e}}1 {í}{{\'i}}1 {ó}{{\'o}}1 {ú}{{\'u}}1
    {Á}{{\'A}}1 {É}{{\'E}}1 {Í}{{\'I}}1 {Ó}{{\'O}}1 {Ú}{{\'U}}1
    {à}{{\`a}}1 {è}{{\`e}}1 {ì}{{\`i}}1 {ò}{{\`o}}1 {ù}{{\`u}}1
    {À}{{\`A}}1 {È}{{\'E}}1 {Ì}{{\`I}}1 {Ò}{{\`O}}1 {Ù}{{\`U}}1
    {ä}{{\"a}}1 {ë}{{\"e}}1 {ï}{{\"i}}1 {ö}{{\"o}}1 {ü}{{\"u}}1
    {Ä}{{\"A}}1 {Ë}{{\"E}}1 {Ï}{{\"I}}1 {Ö}{{\"O}}1 {Ü}{{\"U}}1
    {â}{{\^a}}1 {ê}{{\^e}}1 {î}{{\^i}}1 {ô}{{\^o}}1 {û}{{\^u}}1
    {Â}{{\^A}}1 {Ê}{{\^E}}1 {Î}{{\^I}}1 {Ô}{{\^O}}1 {Û}{{\^U}}1
    {œ}{{\oe}}1 {Œ}{{\OE}}1 {æ}{{\ae}}1 {Æ}{{\AE}}1 {ß}{{\ss}}1
    {ç}{{\c c}}1 {Ç}{{\c C}}1 {ø}{{\o}}1 {å}{{\r a}}1 {Å}{{\r A}}1
    {€}{{\EUR}}1 {£}{{\pounds}}1
    {^}{{{\color{ipython_purple}\^{}}}}1
    {=}{{{\color{ipython_purple}=}}}1
    {+}{{{\color{ipython_purple}+}}}1
    {*}{{{\color{ipython_purple}$^\ast$}}}1
    {/}{{{\color{ipython_purple}/}}}1
    {+=}{{{+=}}}1
    {-=}{{{-=}}}1
    {*=}{{{$^\ast$=}}}1
    {/=}{{{/=}}}1,
    literate=
    *{-}{{{\color{ipython_purple}-}}}1
     {?}{{{\color{ipython_purple}?}}}1,
    identifierstyle=\color{black}\ttfamily,
    commentstyle=\color{ipython_cyan}\ttfamily,
    stringstyle=\color{ipython_red}\ttfamily,
    keepspaces=true,
    showspaces=false,
    showstringspaces=false,
    rulecolor=\color{ipython_frame},
    frame=single,
    frameround={t}{t}{t}{t},
    framexleftmargin=6mm,
    numbers=left,
    numberstyle=\tiny\color{halfgray},
    backgroundcolor=\color{ipython_bg},
    basicstyle=\footnotesize\ttfamily,
    keywordstyle=\color{ipython_green}\ttfamily,
    aboveskip=1.2em,
    belowskip=1.2em,
}

\newcommand{\wcxf}{\texttt{WCxf}}
\newcommand{\json}{\texttt{JSON}}
\newcommand{\yaml}{\texttt{YAML}}
\newcommand{\slha}{\texttt{SLHA}}
\newcommand{\DsixTools}{\texttt{DsixTools}}
\newcommand{\EOS}{\texttt{EOS}}
\newcommand{\flavio}{\texttt{flavio}}
\newcommand{\SARAH}{\texttt{SARAH}\xspace}
\newcommand{\SPheno}{\texttt{SPheno}\xspace}
\newcommand{\FlavorKit}{\texttt{FlavorKit}\xspace}
\newcommand{\FF}{\texttt{FormFlavor}\xspace}
\newcommand{\SMEFTsim}{\texttt{SMEFTsim}\xspace}
\newcommand{\madgraph}{\texttt{MadGraph5\char`_aMC@NLO}}
\newcommand{\SFR}{\texttt{SMEFT Feynman Rules}\xspace}
\newcommand{\wilson}{\texttt{wilson}}

\allowdisplaybreaks

\begin{document}

\begin{frontmatter}

\begin{flushright}
\footnotesize
\vspace*{-3cm}
\begin{tabular}{l}
IFIC/17-61\\
KA-TP-38-2017\\
TUM-HEP-1117/17\\
LMU-ASC 74/17\\
\end{tabular}
\end{flushright}

\begin{center}
\includegraphics[width=3cm]{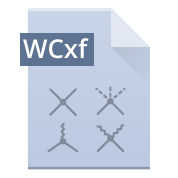}
\end{center}

\title{%
\wcxf:
an exchange format for Wilson coefficients\\
beyond the Standard Model
}

\author[a1]{Jason Aebischer}
\author[nbi]{Ilaria Brivio}
\author[a2]{Alejandro Celis}
\author[a4]{Jared A.~Evans}
\author[nbi]{Yun Jiang}
\author[hri]{Jacky Kumar}
\author[a1]{Xuanyou Pan}
\author[wu]{Werner Porod}
\author[waw]{Janusz Rosiek}
\author[a5]{David Shih}
\author[kit1,kit2]{Florian Staub}
\author[a1]{David M. Straub}
\author[tum]{Danny van Dyk}
\author[a3]{Avelino Vicente}

\address[a1]{Excellence Cluster Universe, TUM, Boltzmannstr.~2, 85748~Garching, Germany}

\address[nbi]{Niels Bohr International Academy and Discovery Center, Niels Bohr Institutet, K\o benhavns Universitet, Blegdamsvej 17, 2100 K\o benhavn \O, Denmark}

\address[a2]{Ludwig-Maximilians-Universit\"at M\"unchen,
   Fakult\"at f\"ur Physik,\\
   Arnold Sommerfeld Center for Theoretical Physics,
   80333 M\"unchen, Germany}

\address[a4]{Department of Physics, University of Cincinnati, Cincinnati, Ohio 45221, USA}

\address[wu]{Institut f\"ur Theoretische Physik und Astrophysik, Universit\"at W\"urzburg, 97074 W\"urzburg, Germany}

\address[waw]{Faculty of Physics, University of Warsaw, Pasteura 5,
  02-093 Warsaw, Poland}

\address[a5]{Department of Physics, Rutgers University, Piscataway, New Jersey 08854, USA}

\address[kit1]{ITP, Karlsruhe Institute of Technology, Engesserstra{\ss}e 7, 76128 Karlsruhe, Germany}
\address[kit2]{IKP, Karlsruhe Institute of Technology, Hermann-von-Helmholtz-Pl.\,1, 76344 Eggenstein-Leopoldshafen, Germany}

\address[tum]{Physik Department, TU M\"unchen, James-Franck-Straße 1, 85748~Garching, Germany}

\address[a3]{Instituto de F\'{\i}sica Corpuscular, Universitat de Val\`encia - CSIC, 46071 Val\`encia, Spain}

\address[hri]{Harish-Chandra Research Institute, Chhatnag Road, Jhusi, Allahabad 211019, India}

\begin{abstract}
We define a data exchange format for numerical values of Wilson coefficients
of local operators parameterising low-energy effects of physics beyond the Standard Model.
The format facilitates interfacing model-specific Wilson coefficient calculators, renormalisation group (RG) runners, and observable calculators.
It is designed to be unambiguous (defining a non-redundant set of operators
with fixed normalisation in each basis), extensible (allowing the addition of new EFTs
or bases by the user), and robust (being based on industry standard file formats with
parsers implemented in many programming languages).
We have implemented the format for
the Standard Model EFT (SMEFT) and for the
weak effective theory (WET) below the electroweak scale
and have added interfaces to a number of public codes dealing with SMEFT or WET.
We also provide command-line utilities and a Python module for convenient
manipulation of \wcxf\ files, including translation between different bases and matching from SMEFT to WET.
\end{abstract}

\end{frontmatter}

\newpage

\section{Introduction and Motivation}\label{sec:intro}

Indirect effects of physics beyond the Standard Model (SM) at energies much lower than the mass scale of the new particles can be described by effective field theories (EFTs) extending the SM by new local operators.
EFT methods can be used either as a means to perform model-independent
analyses of new physics (NP), or merely as a tool to separate the model-independent low-energy phenomenology from the model-dependent short-distance physics when studying specific NP models.

Depending on the problem, different EFTs are appropriate.
For  very low-energy processes like flavour physics, the
renormalisable part of the theory only contains QCD and QED interactions
and light quarks as well as leptons.
For electroweak-scale energies, the Standard Model Effective Field Theory (SMEFT) is appropriate for theories with
a linearly realised breaking of the electroweak symmetry.
Other EFTs relevant for physics beyond the SM include the ``Higgs EFT'' (HEFT) extending the SM
with a non-linear realisation of electroweak symmetry
(see e.g.\ \cite{deFlorian:2016spz,Brivio:2017vri} for reviews on SMEFT and HEFT),
or EFTs with new light degrees of freedom
-- e.g.\ sterile neutrinos or light dark matter particles \cite{Bishara:2017nnn}.
In a phenomenological analysis, typically the Wilson coefficients are
predicted by a NP model at some high scale.
Before predicting experimental observables,
the Wilson coefficients have to be run down to low energies using the
renormalisation group (RG),
and if necessary they have to be matched onto a different EFT appropriate
at lower energies.

There are several public codes dealing with different aspects of EFTs and their corresponding Wilson coefficients, for instance
\begin{itemize}
 \item observable calculators for flavour physics (and beyond), such as \flavio~\cite{flavio}, \EOS~\cite{EOS}, \FlavorKit~\cite{Porod:2014xia,FK-web}, \SPheno~\cite{Porod:2003um,Porod:2011nf}, \FF \cite{Evans:2016lzo,FF-web}, \texttt{SuperIso} \cite{Mahmoudi:2007vz}, \texttt{HEPfit} \cite{Hepfit},
 \item packages for RG evolution or matching, such as \DsixTools{} \cite{Celis:2017hod,DT-web}, \wilson~\cite{Aebischer:2018bkb}, \texttt{MatchingTools} \cite{Criado:2017khh},
 \item packages related to SMEFT, such as \SMEFTsim~\cite{Brivio:2017btx,SMEFTsim-web}, \SFR~\cite{Dedes:2017zog},
 \item codes dark matter EFT, such as \texttt{DirectDM} \cite{Bishara:2017nnn}.
\end{itemize}
A practical hurdle to all these codes is the exchange of numerical values of
Wilson coefficients between them. The challenges include
\begin{itemize}
\item the generally very large number of operators\footnote{%
The Flavour Les Houches Accord (FLHA) \cite {Mahmoudi:2010iz} is a file format for (among other things) the exchange
of numerical Wilson coefficients relevant for flavour physics that solves this problem by defining
a numeric code representing any given operator. While FLHA aims to be a very general exchange format
for flavour physics codes using the WET, the scope of \wcxf{} is different by focusing only
on new physics Wilson coefficients, but also allowing different EFTs and multiple bases.
\wcxf{} Wilson coefficient files in the WET can in principle be converted to FLHA format (but not vice versa)
and we envisage an implementation of this in the future.},
\item the many different bases used in the literature, often geared for a specific problem,
\item ambiguities related to redundant operators or different normalisations,
\item different programming languages used and different quality of parsers for program-specific file formats.
\end{itemize}
The \wcxf{} is an attempt at overcoming these challenges by defining a common file format that is
\begin{itemize}
\item \textit{unambiguous}, by uniquely fixing the set of non-redundant operators and their normalisation in a given basis of a given EFT,
\item \textit{extensible}, by allowing the user to define new EFTs or new bases for existing EFTs,
\item \textit{robust} by using industry standard file formats with parsers implemented in many programming languages.
\end{itemize}
Crucially, the format does not fix a single basis that has to be used by all codes -- acknowledging the fact that different
bases can be more convenient for different problems -- but rather provides a way to \textit{define} a basis (or EFT) by
uploading a definition file to a public repository. The translation between different bases\footnote{%
For translating between different bases, the approach is similar to the Rosetta SMEFT translator \cite{Falkowski:2015wza},
but not restricted to SMEFT.
} (as well as the matching between different EFTs)
can then be performed by a single tool, i.e.\ in principle has to be implemented only once, while individual codes
only have to export or import data using their internal EFT and basis.
Robustness is achieved by using the well-established \yaml{} and \json{} formats.

The rest of this document is organised as follows.
In section~\ref{sec:format}, we define the \wcxf\ format.
In section~\ref{sec:bases}, we discuss the EFTs and bases that we have defined so far.
Section~\ref{sec:cli} describes the command-line tool and Python package that we provide
for the basis translation, matching, and validation of \wcxf\ files.
Section~\ref{sec:codes} contains brief descriptions of the implementation of a \wcxf\
interface in several existing public codes.
Section~\ref{sec:conclusions} contains our conclusions and outlook.

\section{Definition of the format}\label{sec:format}

The \wcxf{} defines three different types of data files:
\begin{itemize}
\item The \textbf{EFT file}, fixing the theory the Wilson coefficients are defined in;
\item The \textbf{basis file}, defining the basis used and listing all
operators defined in the basis;
\item The \textbf{Wilson coefficient file}, containing the actual numerical
values of the Wilson coefficients.
\end{itemize}
The EFT and basis files are meant to be immutable definitions deposited in a public repository,
whereas the Wilson coefficient file contains the actual data to be exchanged between different programs.

The data structure of these files consists of key-value pairs
and lists, but allows for different \textit{metaformats}
to ship this data. The two recommended formats are
\begin{itemize}
 \item \json{}, which can be imported and exported without third-party tools in Python and Mathematica and is very fast to parse due to its rigorous format,
 \item \yaml{}, which is a superset of \json{} (i.e.\ \json{} files can be parsed by
 a \yaml{} parser but not vice versa), but tends to be more human readable/editable and allows the use of comments.
\end{itemize}
In principle, \wcxf\ is not necessarily a \textit{file} format but can also be used to directly
exchange data structures between programs, bypassing the file system
(e.g., in the case of Python this could be a \texttt{wcxf.WC} instance
discussed in section~\ref{sec:python} or simply a dictionary).
The rest of this document will use the \yaml{} format in all examples.

\subsection{The EFT file}

The EFT file only defines the name of the EFT and lists the named \textbf{sectors},
which are sets of operators with definite quantum numbers under global symmetries
preserved by the RG evolution.
In the case of SMEFT \cite{Buchmuller:1985jz},

the RGEs preserve baryon number $B$ and lepton number $L$
separately, so the sectors correspond to transitions with fixed
$\Delta B$ and $\Delta L$. At dimension 5, there is a single
operator with $\Delta L=2$; at dimension 6, there are $B$- and $L$-conserving operators as well as operators with $\Delta B=\Delta L=1$.
A minimal EFT file could thus look like
\begin{lstlisting}[language=yaml]
eft: SMEFT
sectors:
  dB=dL=0:
  dB=dL=1:
  dL=2:
\end{lstlisting}

In the weak effective theory, flavour quantum numbers are conserved by
the QED and QCD RG evolution, such that there are several sectors with
definite flavour quantum numbers, e.g.\ $\Delta S=2$ operators contributing
to neutral kaon mixing.

The rationale of defining sectors is to simplify basis translations and RG
evolution, as they can be performed sector by sector.

\subsection{The basis file}

The basis file contains all the names of the operators grouped by
sectors as defined above.
Operators are not allowed to have flavour indices; i.e., Wilson coefficients
are always scalars.
The basis must not contain redundant operators, but does not necessarily
have to be complete (partial bases may be useful for specific physical problems
or for observable calculators only considering processes sensitive to a subset of all operators).
 A file minimally defining the ``Warsaw basis'' \cite{Grzadkowski:2010es} of SMEFT could look
like
\begin{lstlisting}[language=yaml]
name: Warsaw
eft: SMEFT
sectors:
  dB=dL=0:
    G:
      real: true
    Gtilde:
      real: true
    W:
      real: true
    [...]
    uphi_11:
    uphi_12:
    uphi_13:
    [...]
\end{lstlisting}
The flavour indices (\texttt{11} etc.) are  part of the name. The rationale
for not allowing matrix- or tensor-valued Wilson coefficients is that this would
make it very difficult to define non-redundant bases.
Unless otherwise specified, Wilson coefficients are assumed to be complex-valued.
For Hermitian operators, only allowing real-valued Wilson coefficients,
the key {\YAMLkeystyle real} should be added and set to {\YAMLvaluestyle true}.

As a general convention, the Wilson coefficients are defined as coefficients of
operators in the effective Lagrangian
\begin{equation}
\mathcal L_\text{eff} =
-\mathcal H_\text{eff} =
\sum_{O_i= O_i^\dagger} C_i \, O_i
+
\sum_{O_i\neq O_i^\dagger} \left( C_i \, O_i + C^*_i \, O^\dagger_i\right),
\end{equation}
i.e., the Hermitian conjugate is added only for non-Hermitian operators.
When prefactors are factored out of the Wilson coefficients
(e.g.\ the Fermi constant or CKM elements), these should be understood
as part of the operator.

To define bases that are just a minor modification of an existing basis,
basis files can use \textit{inheritance} by specifying a parent basis name
in the top-level {\YAMLkeystyle parent} key. All sectors omitted from these
child basis files are assumed to be the same\footnote{%
The \textit{name} and \textit{form} of the operators in a sector omitted
from a child basis definition is assumed to be the same as in the parent
basis, but not necessarily the \textit{numerical values}: for instance,
a child basis could inherit all operators from a parent basis, but differ
due to a rephasing of fermion fields.
} as for the parent basis.

\subsection{The Wilson coefficient file}\label{sec:wcf}

While the EFT and basis files only have to be defined once (and are already
pre-defined for a number of standard EFTs and bases), the Wilson
coefficient file contains the actual numerical data to be exchanged between
different codes. Minimally, it just defines the EFT, the basis, the
renormalisation scale, and lists the numerical values of the coefficients.
Unless otherwise specified,
Wilson coefficients are assumed to be renormalised in the
$\overline{\text{MS}}$ scheme.
Dimensionful coefficients must be given in appropriate powers of GeV.
Any coefficient defined in the basis but not present in the data file is
assumed to vanish. If the value is a number,
the Wilson coefficient is assumed to be real.
If it is complex, it must be given as a mapping with keys {\YAMLkeystyle Re} and {\YAMLkeystyle Im}.
Dimensionful numbers have to be given in units of GeV to the appropriate power.
A minimal example could look like
\begin{lstlisting}[language=yaml]
eft: SMEFT
basis: Warsaw
scale: 1e16
values:
    Gtilde: 3.1e-6
    uphi11:
      Re: 0
      Im: 0.0001
\end{lstlisting}

\subsection{Metadata}

While the minimal examples above are sufficient for machine-readable data
exchange, the \wcxf{} format also allows for the addition of metadata.
Metadata can contain textual descriptions of EFTs or bases, information about
the software used to generate the file, or \LaTeX{} code that can be used for
a more readable representation of the basis.

All three file types allow for a top-level \text{metadata} key that can contain
arbitrary key-value pairs below it. Recommended metadata keys include
\begin{itemize}
 \item {\YAMLkeystyle description} in the case of EFT and basis files, containing a
 textual description of the basis and its characteristics,
 \item {\YAMLkeystyle generator} in the case of the Wilson coefficient file, specifying
 the software that generated the file.
\end{itemize}

The basis file, defining the non-redundant set of operators in a given basis,
can additionally contain metadata for each operator.
This is facilitated by the definition in the form of a mapping key.
For instance, the {\YAMLkeystyle tex} key can be used to define the
\LaTeX{} representation of an operator:
\begin{lstlisting}[language=yaml]
    [...]
    H:
      tex: (H^\dagger H)^3
\end{lstlisting}
As a convention, math mode delimiters are omitted.

\section{Pre-defined EFTs and bases}\label{sec:bases}

In this section we describe the EFTs that we have already defined,
namely SMEFT and WET, and the \textit{complete} bases defined therein.
In section~\ref{sec:codes}, we will discuss more specific bases for
WET appropriate for individual codes.

\subsection{SMEFT}\label{sec:smeft}

At dimension 5, SMEFT contains a single operator: the lepton number violating
Weinberg operator generating Majorana masses for the neutrinos~\cite{Weinberg:1979sa}.
At dimension 6, it contains baryon- and lepton number conserving operators
as well as operators with $\Delta B=\Delta L=1$
\cite{Buchmuller:1985jz,Grzadkowski:2010es}.
At dimension 7, there are additional operators with
$\Delta L=2$ as well as with $\Delta B=-\Delta L = 1$ \cite{Lehman:2014jma,Henning:2015alf}.
We limit ourselves to dimension-6 operators for the time being.

The first non-redundant basis for the $\Delta B=\Delta L=0$ operators at dimension
6 was derived in \cite{Grzadkowski:2010es} and for baryon number violating ones
in \cite{Abbott:1980zj}.
This basis is often called the ``Warsaw basis''.
The counting of all non-redundant operators in the presence of three fermion
generations was first performed in \cite{Alonso:2013hga,Alonso:2014zka}.
Recently, a specific choice of a complete set of non-redundant operators
was suggested as part of the \DsixTools{} package \cite{Celis:2017hod}.
We adhere to this choice in our basis file for the Warsaw basis.

A subtlety of choosing a basis for flavoured SMEFT operators is the choice
of the weak basis in the space of the three fermion generations for each of
the quark and lepton fields. Since all fermions are massless in the
$SU(2)_L\times U(1)_Y$-symmetric phase, there is no a priori preferred basis,
such as the mass basis in the weak effective theory below the electroweak scale.
Concretely, the theory is invariant under a $U(3)^5$ rotation,
\begin{align}
\psi &\to U_\psi \psi, &\psi = q, u, d, l, e \,.
\end{align}

While specifying the complete Lagrangian including the Yukawa matrices along with
the Wilson coefficients would fix this basis, this conflicts with the
specification of \wcxf{} as being a Wilson coefficient-only data exchange format.
Instead, our choice for the Warsaw basis is to define a default weak basis
in which the fermion mass matrices have a specific form.
The running fermion mass matrices at the dimension-6 level,
after a general $U(3)^5$ rotation,
can be written as (see e.g.~\cite{Dedes:2017zog})
\begin{align}
M_d &= \frac{v}{\sqrt{2}} \, U_q^\dagger\left(Y_d-\frac{v^2}{2} C_{d\phi}\right)U_d\,,
\label{eq:Md}
\\
M_u &= \frac{v}{\sqrt{2}} \, U_q^\dagger\left(Y_u-\frac{v^2}{2} C_{u\phi}\right)U_u\,,
\label{eq:Mu}
\\
M_e &= \frac{v}{\sqrt{2}} \, U_l^\dagger\left(Y_e-\frac{v^2}{2} C_{e\phi}\right)U_e\,,
\label{eq:Me}
\\
M_\nu &= -v^2 \, U_l^T \, C_{ll\phi\phi} \, U_l \,,
\end{align}
Here $v \approx 246$ GeV is the electroweak vacuum expectation value (VEV)
and $Y_\psi$, with $\psi = u, d, e$, are the SM Yukawa couplings, with
the convention $\displaystyle \mathcal{L}_{\rm SM} \supset - \left(
Y_e \overline{l} e \phi + Y_u \overline{q} u \widetilde{\phi} + Y_d
\overline{q} d \phi \right) \equiv - \sum_\psi Y_\psi
Q_\psi^Y$. $C_{\psi \phi}$ and $C_{ll\phi\phi}$ are the Wilson
coefficients of the SMEFT dimension-6 operators $Q_{\psi \phi} =
\left( \phi^\dagger \phi \right) Q_\psi^Y$ and the lepton number
violating Weinberg operator $Q_{ll\phi\phi} = \left(
\widetilde{\phi}^\dagger l \right)^T C \left( \widetilde{\phi}^\dagger
l \right)$, respectively.
%
We use the freedom of $U(3)^5$ field rotations to choose a basis where
\begin{itemize}
 \item $M_d$ and $M_e$ are diagonal with real positive ascending entries,
 \item $M_u$ has the form $V^\dagger \hat M_u$, where $\hat M_u$ is diagonal
 with real positive ascending entries and $V$ has the form of the CKM matrix
 in standard phase convention \cite{Patrignani:2016xqp}.
 \item $M_\nu$ has the form $U^\ast \hat M_\nu U^\dagger$, where $\hat M_\nu$ is diagonal
 with real positive ascending\footnote{%
 Our definition of $\hat M_\nu$ having \textit{ascending} masses on the diagonal
 means that the matrix $U$, while defined as having the standard phase convention
 of the PMNS matrix, only coincides with the PMNS matrix for a normally ordered
 neutrino mass spectrum. In the case of inverse ordering, $m_{\nu_3}<m_{\nu_1}<m_{\nu_2}$,
 the matrix $U$ corresponds to a PMNS-like matrix with permuted angles.
 } entries and $U$ has the form of the Pontecorvo–Maki–Nakagawa–Sakata (PMNS) neutrino mixing matrix
 in standard phase convention \cite{Patrignani:2016xqp}.
\end{itemize}
This basis choice is convenient as it removes all unphysical parameters present in the
Yukawa couplings and it allows to easily translate to the mass basis at the
scale of electroweak symmetry breaking. However, we stress that this diagonality
is not invariant under SMEFT renormalisation group evolution and only holds at a single
scale\footnote{An additional subtlety is the fact that the VEV $v$ is scale dependent itself
and can even vanish at a high scale. While the overall factor does not affect the
rotation matrices, the $O(v^2)$ terms in \eqref{eq:Md}--\eqref{eq:Me} are affected. To avoid this problem, we advocate using the on-shell definition of $v$
in these terms.}.

As a variant of the default ``Warsaw'' basis, we also define a basis denoted
\texttt{Warsaw up} that uses a weak basis where the up-type quark mass matrix $M_u$
rather than $M_d$ is diagonalised, while $M_d$ has the form $V \hat M_d$.

Finally, we also define a ``Warsaw mass'' basis
coinciding with the basis choice in \cite{Dedes:2017zog} and the $\widetilde C$
basis in \cite{Aebischer:2015fzz}. Due to our weak basis choice in the Warsaw basis, this
basis differs from the ``Warsaw'' basis only by a rotation by the CKM matrix in the operators $O_{uX}$, with $X=\phi, G, W$, or $B$, and by the fact that the neutrino mass operator $O_{ll\phi\phi}$ is diagonal in this basis.

\subsection{Weak effective theory}

The effective theory of the SM below the electroweak scale, in the phase
where the electroweak symmetry is broken, is often called ``weak effective theory''
(WET) as it represents the appropriate EFT for describing weak interactions of
leptons and quarks.\footnote{Alternatively, the WET was dubbed LEFT (low-energy
EFT) in \cite{Jenkins:2017jig}.}
The dynamical fields in WET contain: all leptons; all quarks except the top; the
photon; and the gluon. Since QED and QCD conserve flavour quantum numbers,
the ``sectors'' are simply collections of operators
with fixed flavour quantum numbers that do not mix under RG evolution.
A complete, non-redundant basis and the full 1-loop QED and QCD RG equations
for \textit{flavour-changing operators} were recently presented in \cite{Aebischer:2017gaw}.
The complete basis including flavour-conserving operators and the one-loop anomalous dimension matrix was derived
in \cite{Jenkins:2017jig,Jenkins:2017dyc}.
We provide basis files for both these bases, that we dub ``Bern'' and ``JMS'' bases,
respectively.
For the JMS basis, our basis file defines a complete set of non-redundant operators.
We have checked that the number of non-redundant elements coincides with the
counting in \cite{Jenkins:2017jig}.

For processes below the $b$-quark scale, EFTs with a smaller number of dynamical
quark (and lepton) fields are appropriate. We define four different EFTs
with names and dynamical fields listed in table~\ref{tab:wet}.

Additional WET bases have been defined for specific codes as described in section~\ref{sec:codes}.

\begin{table}[tbp]
\begin{center}
\begin{tabular}{lll}
\toprule
EFT & quarks & leptons \\
\midrule
\texttt{WET} & $u, d, s, c, b$ & $e,\mu,\tau$ \\
\texttt{WET-4} & $u, d, s, c$ & $e,\mu,\tau$ \\
\texttt{WET-3} & $u, d, s$ & $e,\mu$ \\
\texttt{WET-2} & $u, d$ & $e$ \\
\bottomrule
\end{tabular}
\end{center}
\caption{Four EFTs below the electroweak scale and the dynamical quark
and lepton fields they contain.}
\label{tab:wet}
\end{table}

\section{Python and command-line interface}\label{sec:cli}

To facilitate the basis translation and matching of Wilson coefficients
we provide a Python package \texttt{wcxf} including a command-line tool
that can be used by different codes or on its own.
Having this central tool allows in particular to define translation
and matching functions only once and make them available to different
programs.
We have already implemented the matching from SMEFT (in the Warsaw basis) to
the WET (in the JMS basis) as well as translations between most of the bases
discussed here.
The tool also allows validating the format of \wcxf\ files.
Here
we briefly describe its main features and refer to the project web site
\cite{wcxf-web}
for detailed documentation.

\subsection{Installation}

The Python package and command line interface require Python version 3.5 or above as well as the \texttt{NumPy} package. With these prerequisites, they can be installed by the command\footnote{The name of the Python 3 executable might differ depending on the system.}
\begin{lstlisting}[language=iPython]
python3 -m pip install wcxf --user
\end{lstlisting}
This will also download the EFT and basis files from the public repository.
When a new version is available, the package can be upgraded with
\begin{lstlisting}[language=iPython]
python3 -m pip install --upgrade wcxf --user
\end{lstlisting}
A development version of \wcxf\ can also be installed directly from a Git working directory using
\begin{lstlisting}[language=iPython]
python3 -m pip install --user -e .
\end{lstlisting}

\subsection{Python package}\label{sec:python}

The \texttt{wcxf} package provides three classes representing EFT, basis, and Wilson coefficient files, aptly named \texttt{wcxf.EFT}, \texttt{wcxf.Basis}, and \texttt{wcxf.WC}. All three classes provide a \texttt{load} method to
load \yaml\ or \json\ files as well as a \texttt{validate} method that
performs various checks on the file and raises an exception if a problem
is found. For instance, a Wilson coefficient file can be read and validated
via
\begin{lstlisting}[language=iPython]
import wcxf

with open('my_wcxf_input_file.yml', 'r') as f:
  wc = wcxf.WC.load(f)
wc.validate()
\end{lstlisting}
The \texttt{WC} class also has a \texttt{dict} property that returns the numerical values of all Wilson coefficients as a dictionary, already representing complex coefficients by complex numbers rather than a dictionary of real and imaginary parts.

All EFTs and bases present in the public repository are part of the package and are automatically read in at import time. Existing EFTs and bases can be accessed by their names as follows,
\begin{lstlisting}[language=iPython]
smeft = wcxf.EFT['SMEFT']
warsaw_basis = wcxf.Basis['SMEFT', 'Warsaw'] # eft, basis
\end{lstlisting}

Translators between different bases of the same EFT can be defined by writing a function accepting a dictionary of Wilson coefficient name-value pairs in the source basis, and returning a corresponding dictionary in the target basis. The translator can then be made known to the package simply by decorating it with the \texttt{translator} decorator,
\begin{lstlisting}[language=iPython]
@wcxf.translator('My source basis', 'My target basis')
def my_translation_function(source_dict):
  # ...
  return target_dict
\end{lstlisting}
Matchers from one EFT to another are defined analogously by a \texttt{matcher} decorator, where both the EFT and the basis of the source and target dictionaries have to be specified,
\begin{lstlisting}[language=iPython]
@wcxf.matcher('My UV EFT', 'My UV EFT basis', 'My IR EFT', 'My IR EFT basis')
def my_matching_function(source_dict):
  # ...
  return target_dict
\end{lstlisting}

If the appropriate translator or matcher exists, a Wilson coefficient instance in the new basis or EFT can be generated by calling the \texttt{translate} or \texttt{match} methods with the source basis (or EFT and basis) as arguments,
\begin{lstlisting}[language=iPython]
wc_new = wc_old.translate('My target basis')
wc_new = wc_old.match('My IR EFT', 'My IR EFT basis')
\end{lstlisting}

\subsection{Command-line interface}

The validation, translation, and matching functionality of the Python package are
also available through a command-line script that is automatically installed along
with the package.
In addition, the script can also convert \yaml\ files to \json\ and vice versa.
The script is invoked as\footnote{%
If the command is not found even though the package has been installed,
on Linux you might have to add \texttt{\$HOME/.local/bin} to your \texttt{\$PATH}.
}
\begin{lstlisting}[language=iPython]
wcxf <command> [<options>]
\end{lstlisting}
where \texttt{<command>} can be
\begin{itemize}
 \item \texttt{convert}  for converting between the \json\ and \yaml\ formats,
 \item \texttt{validate}  for validating that a file adheres to the standard,
 \item \texttt{translate}  for translating between different bases,
 \item \texttt{match}  for matching between different EFTs.
\end{itemize}
The available options and arguments can be displayed by calling
\begin{lstlisting}[language=iPython]
wcxf <command> -h
\end{lstlisting}
Here we just list a few common examples,
\begin{lstlisting}[language=iPython]
wcxf convert json my_file.yml --output my_file.json      # YAML -> JSON conversion
wcxf validate basis my_basis.yml                         # Basis validation
wcxf validate wc my_coeffs.yml                           # Wilson c. validation
wcxf translate flavio wc_jms.yml --output wc_flavio.json # Basis translation
wcxf match WET JMS wc_warsaw.json --output wc_jms.json   # SMEFT -> WET matching
\end{lstlisting}

\section{Implementation in public codes}\label{sec:codes}

\subsection{\flavio}

\flavio\ \cite{flavio} is a Python package for flavour physics phenomenology in the SM and beyond.
The features include making predictions for a host of flavour observables in the presence
of new physics parameterised by Wilson coefficients of dimension-6 operators in the WET,
and fitting Wilson coefficients to experimental data using Bayesian or frequentist methods.
The interface for setting values of new physics Wilson coefficients in \flavio\
is the \texttt{WilsonCoefficient} class. From version 0.25, this class
supports loading the initial values of the Wilson coefficients from a \wcxf\
file, making use of the \texttt{wcxf-python} package (cf.\ section~\ref{sec:python}). A simple example how to read the Wilson coefficients from a file:
\\\noindent
\begin{minipage}{\linewidth}
\begin{lstlisting}[language=iPython]
import flavio, wcxf

with open('my_wcxf_input_file.yml', 'r') as f:
  wc = wcxf.WC.load(f)

fwc = flavio.WilsonCoefficients()
fwc.set_initial_wcxf(wc)
\end{lstlisting}
\end{minipage}
The initial values are set at the scale specified in the \wcxf\ file;
\flavio\ takes care of the RG evolution to the scale relevant for each
process automatically. The EFT of the input must be WET, but the basis
is arbitrary, as long as a translator to the \flavio\ basis is defined
in the \texttt{wcxf-python} package or the user's script.
From version 0.28, \flavio{} alternatively supports using 
the \texttt{Wilson} class from the \wilson{} package
as a drop-in replacement for \texttt{flavio.WilsonCoefficients},
further simplifying the workflow (see section~\ref{sec:wilson} for a description).

\subsection{\EOS}

\EOS\ is a software library written in C++14 that fulfills two use cases: first, it
can be used for the computation of flavour physics observables within the SM or beyond.
Beyond the SM physics is implemented through a WET with operators up to dimension six.
Second, \EOS\ allows to infer knowledge of SM and WET parameters from experimental constraints
within a Bayesian framework. \EOS\ provides command-line clients for the most common
tasks, as well as library interfaces in both C++ and Python; see \cite{EOS}
for an introduction and a user manual.

The command-line script \texttt{wcxf2eos}, that is shipped with the \texttt{wcxf}
Python package, converts a \wcxf\ file to a \EOS\ parameter \yaml\ file
\begin{lstlisting}[language=iPython]
wcxf2eos my_wcxf_input_file.yml --output eos.yaml
\end{lstlisting}
The use of this parameter file is most convenient when examining a single point
in the WET parameter space. For examining more than a few points, we recommend
using the \EOS\ Python interface instead:
\begin{lstlisting}[language=iPython]
import eos, wcxf

with open ('my_wcxf_input_file.yml', 'r') as f:
    wc = wcxf.WC.load(f)

parameters = eos.Parameters.FromWCxf(wc)
\end{lstlisting}
At the time of writing this document, \EOS\ implements $b\to s\lbrace qq, \ell\ell, \gamma, g\rbrace$
as well as $b\to \lbrace c,u\rbrace \ell\nu$ transitions in a variety of exclusive and inclusive decay modes.
As of version 0.2, violation of lepton flavour universality is implemented for $b\to s\ell\ell$ and
$b\to \lbrace c, u\rbrace \ell\nu$ transitions.

\subsection{\FlavorKit}
\label{sec:flavorkit}
\FlavorKit \cite{Porod:2014xia,FK-web} is an extension of the {\tt
  Mathematica} package \SARAH
\cite{Staub:2008uz,Staub:2009bi,Staub:2010jh,Staub:2012pb,Staub:2013tta}
that increases its capability to handle flavour observables.  Since it
is based on \SARAH, \FlavorKit is not restricted to a specific model,
but can be used to obtain analytical and numerical predictions for
quark and lepton flavour observables in a wide range of models.
\FlavorKit adds to the \SPheno \cite{Porod:2003um,Porod:2011nf}
interface of \SARAH all necessary routines for the numerical
calculation of Wilson coefficients at the full one-loop level in a
given theory. In previous versions of \FlavorKit, the numerical values
of these Wilson coefficients were written by \SPheno in a standard
output file in FLHA format. With version 4.12.3 of \SARAH, the support
of the \wcxf{} format has also been added.

\SPheno will export the numerical values of all calculated Wilson
coefficients into \json{} output files following the \wcxf{} format if
the flag
\begin{lstlisting}[language=iPython]
Block SPhenoInput #
...
79 1    # Write WCXF files (exchange format for Wilson coefficients)
\end{lstlisting}
is set in the Les Houches input file. The generated files are called
{\tt WC.\$MODEL\char`_\$X.json}, where {\tt \$MODEL} is the name of the
considered model and {\tt \$X} counts the scales for which the
coefficients are written out. \SPheno computes the Wilson coefficients
at two different scales: the coefficients for quark flavour violating
operators are calculated at $Q=160$~GeV whereas those for lepton
flavour violating operators are calculated at $Q=91$~GeV.\footnote{The
  internal classification of the effective operators into quark and
  lepton flavour violating is explicitly given in Appendix A of the
  \FlavorKit manual \cite{Porod:2014xia}. We note that some of the
  quark flavour violating operators violate lepton flavour as well.}
Therefore, two \json{} files are written by \SPheno:
\begin{itemize}
 \item {\tt WC.\$MODEL\char`_1.json}: this contains the Wilson coefficients for quark flavour violating operators calculated at $Q=160$~GeV
 \item {\tt WC.\$MODEL\char`_2.json}: this contains the Wilson coefficients for lepton flavour violating operators calculated at $Q=91$~GeV
\end{itemize}
These output files contain the numerical results for the Wilson
coefficients in the \FlavorKit basis~\cite{Porod:2014xia}. This WET
basis consists of a non-redundant set of operators for quark and
lepton flavour physics. In contrast to the JMS basis
\cite{Jenkins:2017jig}, which is very similar, the \FlavorKit basis is
not complete and can be further extended with the addition of new
operators. This can be done by means of the {\tt PreSARAH} package,
which interfaces \FlavorKit with {\tt FeynArts}/{\tt
  FormCalc}~\cite{Hahn:1998yk,Hahn:2000kx,Hahn:2000jm,Hahn:2004rf,Hahn:2005vh,Nejad:2013ina}
to fully automate the one-loop calculation of the Wilson
coefficients. We refer to the \FlavorKit manual \cite{Porod:2014xia}
for more details. Once the new operators are added with {\tt
  PreSARAH}, \FlavorKit must be adapted to generate output \wcxf{} files
containing the new Wilson coefficients. This is achieved by modifying
the file
\begin{center}
{\tt SARAH-X.Y.Z/FlavorKit/WCXF\char`_WilsonCoefficients.m}
\end{center}
This modification must be applied before generating the \SPheno code for the model of
interest. The standard format of the {\tt WCXF\char`_WilsonCoefficients.m}
file is the following:
\\\noindent
\begin{minipage}{\linewidth}
\begin{lstlisting}[language=MyMathematica]
WCXF`Outputs = 2;

(* Output for QFV at Q=160 GeV *)
WCXF`EFT[1]= "WET";
WCXF`Basis[1]= "FlavorKit";
WCXF`Scale[1]= 160.0;

WCXF`Values[1]={

(* SECTOR: SBSB *)
(* 4D *)
{{"DVLL_2323",O4dVLL[2,3,2,3], Complex}},

....

(* SECTOR: UBENU *)
(* UDENU *)
Table[{"GVLL_"<>ToString[3]<>ToString[1]<>ToString[i]<>ToString[1],OdulvVLL[3,1,i,1], Complex},{i,1,3}],

....

};
WCXF`Values[1]=Flatten[WCXF`Values[1],1];

(* Output for LFV at Q=91 GeV *)
WCXF`EFT[2]= "WET";
WCXF`Basis[2]= "FlavorKit";
WCXF`Scale[2]= 91.0;

WCXF`Values[2]={

....

};
WCXF`Values[2]=Flatten[WCXF`Values[2],1];
\end{lstlisting}
\end{minipage}
First, {\tt WCXF`Outputs} is used to define the number of energy
scales for which an output \json{} file shall be written. For each of
these scales, the EFT ({\tt WCXF`EFT}), the basis ({\tt WCXF`Basis})
and the value of energy scale in GeV ({\tt WCXF`Scale}) are
set. Finally, the Wilson coefficients in this basis are related to the
coefficients internally calculated by \FlavorKit. The general syntax
for each coefficient is:
\begin{lstlisting}[language=MyMathematica]
{Name WCXF, Name FlavorKit, Complex or Real}
\end{lstlisting}
One should note that while the name of the coefficient used in the
\wcxf{} files is a string, the \FlavorKit name is a {\tt Mathematica}
symbol. Furthermore, the conventions for the \FlavorKit name described
in Appendices A and B of \cite{Porod:2014xia} must be taken into
account.  For instance, in the example given above, the first Wilson
coefficient corresponds to the operator $(\bar b \gamma^\mu P_L
s)(\bar b \gamma_\mu P_L s)$.  Since the coefficients are usually
defined for all three generations of fermions, it might be helpful to
use {\tt Mathematica} functions like {\tt Table} to define them in a
compact form. This is also shown in the example, where the Wilson
coefficients of the $(\bar u \gamma^\mu P_L b)(\bar e \gamma_\mu P_L
\nu_i)$ operators, with $i=1,2,3$, are also defined.

\subsection{\SPheno}
The stand-alone \SPheno package \cite{Porod:2003um,Porod:2011nf} has included the \wcxf{} format
starting with version 4.1.0  in a similar manner  as  the
\FlavorKit package \cite{Porod:2014xia,FK-web} described above.
\SPheno will export the numerical values of all calculated Wilson
coefficients into \json{} output files following the \wcxf{} format if
the flag
\begin{lstlisting}[language=iPython]
Block SPhenoInput #
...
79 1    # Write WCXF files (exchange format for Wilson coefficients)
\end{lstlisting}
is set in the Les Houches input file. The generated files are called
{\tt WC.SPheno\char`_\$X.json}, where  {\tt \$X} counts the scales for which the
coefficients are written out. \SPheno computes the Wilson coefficients
at two different scales: the coefficients for quark flavour violating
operators are calculated at $Q=160$~GeV whereas those for lepton
flavour violating operators are calculated at $Q=m_Z$.
Therefore, two \json{} files are written by \SPheno:
\begin{itemize}
 \item {\tt WC.SPheno\char`_1.json}: this contains the Wilson coefficients for quark flavour violating operators calculated at $Q=160$~GeV
 \item {\tt WC.SPheno\char`_2.json}: this contains the Wilson coefficients for lepton flavour violating operators calculated at $Q=m_Z$
\end{itemize}
These output files contain the numerical results for the Wilson
coefficients in the \FlavorKit basis~\cite{Porod:2014xia} discussed in section \ref{sec:flavorkit}.
 However, only those
coefficients are calculated  which are needed for the flavour observables
calculated in \SPheno, see  \cite{Porod:2011nf} for the corresponding list.

\subsection{\DsixTools}

\DsixTools{} \cite{Celis:2017hod,DT-web} is a Mathematica package for
the handling of the SMEFT. It includes modules for the one-loop RG
evolution from the NP scale down to low energies \cite{Jenkins:2013zja,Jenkins:2013wua,Alonso:2013hga}, tree-level
matching to the WET at the electroweak scale \cite{Aebischer:2015fzz} and RG evolution of
the WET coefficients down to low energies \cite{Aebischer:2017gaw}.

Since version 1.1.2, \DsixTools{} supports the handling of input and
output files for the Wilson coefficients in \wcxf{} format. Regarding
the input, this is accomplished by extending the functionality of the usual \DsixTools{} routine to read input cards,
\texttt{ReadInputFiles}. The routine can be used as
\begin{lstlisting}[language=MyMathematica]
ReadInputFiles["Options.dat", "WCsInput.json", "SMInput.dat"];
\end{lstlisting}
The \wcxf{} input file can be provided in \json{} or \yaml{} formats. Note however that reading input files in \yaml{} format requires previous installation of a \yaml{} importer for Mathematica~\cite{myaml}. This file should contain the Wilson
coefficients in the Warsaw basis as defined in
section~\ref{sec:smeft}, and it must be supplemented with the files
{\tt Options.dat} and {\tt SMInput.dat}, containing the working
options and SM parameters, respectively.  Note that in this case the
SM parameters must be supplied in the weak basis defined in
section~\ref{sec:smeft}. Internally, this basis is known in
\DsixTools{} as \textit{WCXF basis} and is different from the
\textit{fermion mass basis}.\footnote{One should also keep in mind
  that Wilson coefficients are dimensionless in \DsixTools{}, which
  implies $C_{\rm \tt
    WCxf} = \frac{1}{\Lambda^2} \, C_{\rm \tt DsixTools}$, with $\Lambda$ the high scale at which the Wilson
  coefficients are generated.} \texttt{ReadInputFiles} first
translates the \wcxf{} input file to \slha{} format, the default
format in \DsixTools{}, inspired by the Supersymmetry Les Houches
Accord~\cite{Skands:2003cj,Allanach:2008qq}, and then proceeds as
usual. As a result of this, the file {\tt WCsInput.dat}, with the
resulting Wilson coefficients in \slha{} format, is also produced when
executing \texttt{ReadInputFiles}. Additionally, the user can also
use \DsixTools{} to convert a \wcxf{} file to \slha{} format simply as
\begin{lstlisting}[language=MyMathematica]
WCXFtoSLHA["WCs.json", "WCs.dat", HIGHSCALE];
\end{lstlisting}
where in this case {\tt WCs.json} and {\tt WCs.dat} are the input \wcxf{} and
output \slha{} files, respectively, and the high scale $\Lambda$, at which the coefficients are generated, is given in GeV in the argument {\tt HIGHSCALE}. \DsixTools{} can also export the
SMEFT Wilson coefficients in \wcxf{} format. First, the user must load
the {\tt EWmatcher} module of \DsixTools{}, where all the required
routines are contained. This is done with
\begin{lstlisting}[language=MyMathematica]
LoadModule["EWmatcher"]
\end{lstlisting}
Once this module is loaded, one must rotate the SMEFT Wilson
coefficients to the WCXF basis, ensuring that the standard CKM phase
conventions are followed, see section~\ref{sec:smeft}. This can be
done by executing the routine
\begin{lstlisting}[language=MyMathematica]
RotateToWCXFBasis;
\end{lstlisting}
This routine creates the replacement rule {\tt ToWCXFBasis}, which can
then be used to obtain the values of specific Wilson coefficients in the
WCXF basis. For instance,
\begin{lstlisting}[language=MyMathematica]
LQ1[2, 2, 2, 3] /. ToWCXFBasis
\end{lstlisting}
would give the numerical value of the $\left( C_{\ell q}^{(1)}
\right)_{2223}$ coefficient in this basis. One can proceed analogously
with any SMEFT Wilson coefficient. After running {\tt
  RotateToWCXFBasis}, the user can export all the Wilson coefficients
in the WCXF basis to a \wcxf{} file with
the command
\begin{lstlisting}[language=MyMathematica]
WriteWCsOutputFile["WCs.json", "WCXF", "JSON"];
\end{lstlisting}
or similarly
\begin{lstlisting}[language=MyMathematica]
WriteWCsOutputFile["WCs.yaml", "WCXF", "YAML"];
\end{lstlisting}
where the second argument of the routine indicates the fermion basis
and the third the output file format. Finally, one can also convert a
Wilson coefficients file in \slha{} format to \wcxf{} format using
\begin{lstlisting}[language=MyMathematica]
SLHAtoWCXF["WCs.dat", "WCs.json", CPV, SCALE, HIGHSCALE];
\end{lstlisting}
Note that the \slha{} file does not contain information about whether
CP violating entries are allowed in the Wilson coefficients, about the
value of the scale at which these are given ({\tt SCALE}) or about the
high scale ({\tt HIGHSCALE}) at which they are generated. This is why
the user must provide these three details when running this routine.

\subsection{\texttt{wilson}}\label{sec:wilson}

\wilson~\cite{Aebischer:2018bkb} is a Python library for matching and running Wilson coefficients of
higher-dimensional operators beyond the Standard Model. It includes the complete 1-loop SMEFT RGE running, the full tree-level matching onto the WET as well as the complete 1-loop WET (QCD and QED) running. \wilson{} allows the user to specify Wilson coefficients at the UV scale and extract them at any lower scale, possibly below the EW scale. The running and matching procedure is done automatically by \wilson. To define an initial set of Wilson coefficients in \wilson{} one has to specify its name, value,  scale, EFT, and basis. As an example, a parameter point with the Wilson coefficient $[C_{uG}]_{33}$ equal to $1/\text{TeV}^2$ at the scale 1~TeV in the Warsaw basis in SMEFT would be defined as
\begin{lstlisting}[language=iPython]
from wilson import Wilson
mywilson = Wilson({'uG_33': 1e-6},
                  scale=1e3, eft='SMEFT', basis='Warsaw')
\end{lstlisting}
Since \wilson{} is based on the \wcxf{} Python API (see section~\ref{sec:python}), 
the initial values can be simply returned as an instance of \texttt{wcxf.WC} with
\begin{lstlisting}[language=iPython]
mywilson.wc
\end{lstlisting}
Furthermore one can initialize the \texttt{WC} class directly from a file
in \wcxf{} format via
\begin{lstlisting}[language=iPython]
from wilson import Wilson
with open('my_wcxf_input_file.json') as f:
    mywilson = Wilson.load_wc(f)
\end{lstlisting}

\subsection{\FF}

\newcommand{\FA}{\texttt{FeynArts}\xspace}
\newcommand{\FC}{\texttt{FormCalc}\xspace}
\newcommand{\CA}{\texttt{CalcAmps}\xspace}
\newcommand{\FFM}{\texttt{FFModel}\xspace}
\newcommand{\FFW}{\texttt{FFWilson}\xspace}
\newcommand{\FFO}{\texttt{FFObservables}\xspace}

\FF \cite{Evans:2016lzo,FF-web} is a {\tt Mathematica} package for
deriving Wilson coefficients from scratch in a general new physics model.
The code is modular across several pieces: \CA for deriving analytic
expressions for Wilson coefficients with the aid of the packages \FA
\cite{Hahn:2000kx} and \FC \cite{Hahn:1998yk}; \FFW for converting
analytic expressions into numerical Wilson coefficients; \FFO for
converting numerical Wilson coefficients into flavour observables; and
\FFM, containing model specific files that interface with \CA and \FFW.

As of version 1.2.0, \FF has the ability to read and write \wcxf\ files.
The output of \FFW (and associated routines such as
{\tt FFWilsonfromSLHA2}) can be written to a \json\ file with the
command
\begin{lstlisting}[language=MyMathematica]
WriteWilsonToJSON[WILSON, "path/filename.json"];
\end{lstlisting}
where {\tt WILSON} is the output of \FFW and {\tt path/filename.json}
is the desired output file and path.  All Wilson coefficients are
output in the \FF basis \cite{Evans:2016lzo} at the scale (normally,
a high scale) where the coefficients have been evaluated.  Information
contained in the \FFW output on the origin of the contribution (i.e.,
which diagrams provide the information) is lost in this output.   \json\
files can also be read for use with \FFO by the command
\begin{lstlisting}[language=MyMathematica]
ReadWilsonFromJSON["path/filename.json"];
\end{lstlisting}
where the input file must be in the {\tt formflavor} basis.

\subsection{\SMEFTsim}
\SMEFTsim~\cite{Brivio:2017btx,SMEFTsim-web} is a package containing a set of FeynRules~\cite{Christensen:2008py} and UFO~\cite{Degrande:2011ua} models implementing the complete set of dimension-six operators in the Warsaw basis that conserve $B$ and $L$. It allows symbolical calculations in Mathematica as well as numerical simulations e.g. in \madgraph~\cite{Alwall:2014hca}. \SMEFTsim contains 6 different models, corresponding to 3 possible flavour assumptions (a flavour general case, a $U(3)^5$ symmetric limit and a linear MFV case\footnote{This case is based on the MFV paradigm~\cite{DAmbrosio:2002vsn}, that assumes that the only source of CP violation in the $d\leq6$ Lagrangian is the CP-odd phase of the CKM matrix and that a $U(3)^5$ flavour symmetry is present, broken only by insertions of the Yukawa couplings. The automated implementation contains all the relevant flavour spurion insertions up to linear order in $(Y_fY_f^\dagger),\, (Y_f^\dagger Y_f)$.}) and 2 input parameter scheme choices,
$\{\alpha_{ew}, m_Z, G_F\}$ and $\{m_W, m_Z, G_F\}$.

The basis used in \SMEFTsim coincides with the ``Warsaw mass'' basis predefined in \wcxf, up to small notational differences: all the Wilson coefficients are adimensional and a cutoff scale $\Lambda$ is defined as an independent parameter with a default value of 1~TeV. Moreover, complex Wilson coefficients are parameterised defining their absolute value and complex phase instead of the real and imaginary parts.

The \wcxf\ Python package contains a command-line script \texttt{wcxf2smeftsim}, allowing to convert a \json\ or \yaml\ file into a \texttt{param\char`_card.dat}, that can be used to run numerical simulations in \madgraph\ with any of the flavor-general \SMEFTsim models. The script allows to specify a few optional parameters: the
desired input scheme (which can be either \texttt{alpha} or \texttt{mw}) and \SMEFTsim model set (A or B), the name of the output file and the
value to be assigned to the cutoff scale $\Lambda$, that can be an arbitrary number in GeV. For instance:
\definecolor{ipython_green}{RGB}{0, 0, 0}
\begin{lstlisting}[language=iPython]
   wcxf2smeftsim my_wcxf_input_file.json --output my_param_card.dat \
   --model-set A --input-scheme alpha --cutoff-scale 1000
\end{lstlisting}
\definecolor{ipython_green}{RGB}{0, 128, 0}
If \texttt{model-set}, \texttt{input-scheme} or \texttt{cutoff-scale} are not specified, they are assigned to the default values \texttt{A}, \texttt{alpha} and 1000 respectively. By default the output file is called \texttt{wcxf2smeftsim\_param\_card.dat}.

\subsection{\SFR}

\SFR~\cite{Dedes:2017zog} is a Mathematica package evaluating the
Feynman rules for the SMEFT in terms of the physical
(mass-eigenstates) fields of the theory. It works using the {\tt
  FeynRules} package~\cite{Alloul:2013bka}.  Depending on the users'
choice, the package is able to calculate Feynman rules analytically
for the unitary or for a general $R_\xi$ gauge (including ghost
interactions), for the full model or for a chosen field sector and/or
subset of higher dimension operators.  The results are available in
Mathematica or \LaTeX{} formats. Using {\tt FeynRules} interfaces they
can also be exported to formats accepted by other symbolic packages
(e.g.\ {\tt FeynCalc}) or to the UFO format~\cite{Degrande:2011ua} used by
Monte Carlo generators like MadGraph.

As the \SFR{} package is designed to produce the analytical form of
the SMEFT Lagrangian and interaction vertices in terms of the physical
fields, the package itself does not need the numerical values of the Wilson
coefficients. However, the derived Feynman rules are just the
intermediate step in the calculations of physical observables and
usually are fed as input to other programs, calculating such
quantities and using the numerical input for Wilson coefficients.
Thus, starting from version 1.1, \SFR{} is able to import and export
values of the Wilson coefficients in \wcxf{} format.

The import routine reads the \wcxf{} input from the Wilson coefficient
file in the \json{} format (see Sec.~\ref{sec:wcf}) and produces a model
file containing the SMEFT parameters in the {\tt FeynRules} format. It
assumes that the imported values of Wilson coefficients are defined in
the ``Warsaw mass'' basis defined at the end of
Section~\ref{sec:smeft} (see also Refs.~\cite{Dedes:2017zog,
  Aebischer:2015fzz}).  The transcription from the \wcxf{} format can
be done using the following command (in the Mathematica notebook):
\begin{lstlisting}[language=MyMathematica]
<< smeft_wcxf.m

WCXFtoSMEFT["wcxf_input_file", "smeft_par_model_file" ];
\end{lstlisting}
The {\tt `smeft\char`_par\char`_model\char`_file'} is generated in the {\tt
  `definitions'} sub-directory of the \SFR{} package tree.  Next, the
user should edit the {\tt `smeft\char`_control\char`_variables.m'} file
accordingly, defining the correct name of the model for the file for
the UFO format generator:
\begin{minipage}{\linewidth}
\begin{lstlisting}[language=MyMathematica]
...
$UFOModelFile = "smeft_par_model_file";
...
\end{lstlisting}
\end{minipage}
In order to use this model file, the user should first run the {\tt
  `smeft\char`_initialize.m'} program (see program web page \cite{SFRweb} for more detailed instructions):
\begin{lstlisting}[language=MyMathematica]
<< smeft_initialize.m;
\end{lstlisting}
which calculates the analytical form of physical SMEFT Lagrangian,
without assigning numerical values to the Wilson coefficients, and
stores it to disk files. Next, in the new Mathematica notebook (the
{\tt FeynRules} package cannot reload model file without being
restarted), the user should run
\begin{lstlisting}[language=MyMathematica]
<< smeft_UFO.m;
\end{lstlisting}
The {\tt `smeft\char`_UFO.m'} program reads the physical SMEFT Lagrangian
with the Wilson coefficients initialised to the numerical values
defined in the model file generated from \wcxf{} input, generates the
Feynman rules in the form of the UFO output and, if necessary, could
also be adapted to produce other types of output formats supported by
{\tt FeynRules}.  {\tt `smeft\char`_UFO.m'} can be executed multiple times
(always in new Mathematica notebooks), after each modification of the
Wilson coefficients values, without rerunning the {\tt
  `smeft\char`_initialize.m'} program.

The inverse translation from {\tt FeynRules} to \wcxf{} \json{} format
can be done using the commands (files should be in the current
directory or full paths must be specified in their names):
\begin{lstlisting}[language=MyMathematica]
<< smeft_wcxf.m

SMEFTtoWCXF["smeft_par_model_file", "wcxf_output_file"];
\end{lstlisting}

\section{Conclusions and outlook}\label{sec:conclusions}

We have defined a new data exchange format for Wilson coefficients of local operators
parameterising low-energy effects of physics beyond the Standard Model. By implementing this
format both for the Standard Model Effective Field Theory (SMEFT) above and the Weak Effective Theory
(WET) below the electroweak scale in several public codes, we have demonstrated that this format
facilitates interfacing different codes dealing with precision tests of the Standard Model.
This makes it more realistic to construct global likelihoods -- including flavour physics,
electroweak precision tests, and Higgs physics -- for model-independent analyses of new physics
in the future. Additionally, model-specific analyses can be greatly simplified, since the EFT
can serve as as a tool to separate the model-independent low-energy phenomenology
from the model-dependent short-distance physics. This can even be automatised, as demonstrated
by \texttt{FlavorKit} and \texttt{FormFlavor} for WET, and more recently by a complete
tree-level matching in SMEFT \cite{Criado:2017khh,deBlas:2017xtg}.

Since the format is extensible, we encourage the submission of new bases or EFTs to the
public repository\footnote{\url{https://github.com/wcxf/wcxf-bases}} via a pull request.
The documentation of the bases and associated software is available on the project website.\footnote{%
\url{https://wcxf.github.io}}

\section*{Acknowledgements}

\noindent
The work of D.\,S., J.\,A., and X.\,P.\ is supported by the DFG cluster of excellence ``Origin and Structure of the Universe''.  The work of A.\,C. is supported by the DFG grant BU 1391/2-1. I.\,B. and Y.\,J. are supported by the Villum Foundation, NBIA, the Discovery Centre at Copenhagen University and the Danish National Research Foundation (DNRF91).  A.\,V. acknowledges financial support from the ``Juan de la Cierva'' program (27-13-463B-731) funded by the Spanish MINECO as well as from the grants FPA2014-58183-P, FPA2017-85216-P and SEV-2014-0398 (MINECO), and PROMETEOII/ 2014/084 (Generalitat Valenciana). D.\,v.\,D. is supported by the Deutsche Forschungsgemeinschaft (DFG) within the Emmy Noether programme under grant DY 130/1-1 and through the DFG Collaborative Research Center 110 ``Symmetries and the Emergence of Structure in QCD''.
The work of J.R. is supported in part by the National Science Centre,
Poland, under research grants DEC-2015/19/B/ST2/02848 and
DEC-2015/18/M/ST2/00054. W.\,P.\ is supported by the DFG, project nr.\ PO 1337-7/1.
F.\,S.\ is supported by the ERC Recognition Award ERC-RA-0008 of the Helmholtz Association.
I.\,B. and Y.\,J. also thank Michael Trott for valuable discussions and suggestions in designing the \SMEFTsim interface. A.\,C. and A.\,V. thank Javier Fuentes-Mart\'in and Javier Virto for their collaboration in the development of \DsixTools.
D.\,S.\ thanks Alex Arbey, Christoph Bobeth, Nazila Mahmoudi, Ayan Paul, and Mauro Valli for useful discussions.

\section*{References}

\bibliography{bibliography}

\end{document}